Title Page

# Contact inhibition of locomotion and mechanical cross-talk between cell-cell and cell-substrate adhesion determines the pattern of junctional tension in epithelial cell aggregates.

Luke Coburn[1,2*], Hender Lopez[1,3] Benjamin J. Caldwell[5], Elliott Moussa[5], Chloe Yap[5], Rashmi Priya[5], Adrian Noppe[4], Anthony P. Roberts[4], Vladimir Lobaskin[1], Alpha S. Yap[5], Zoltan Neufeld[4,5], Guillermo A. Gomez[5*]

Affiliations: [1]School of Physics and Complex and Adaptive Systems Laboratory, University College Dublin, Dublin, Ireland; [2]Institute of Complex Systems and Mathematical Biology, University of Aberdeen, United Kingdom; [3]Center for BioNano Interactions, School of Chemistry and Chemical Biology, University College Dublin, Belfield, Ireland, [4]School of Mathematics and Physics and [5]Institute for Molecular Bioscience, Division of Cell Biology and Molecular Medicine, The University of Queensland, St. Lucia, Brisbane, Queensland, Australia 4072

*To whom correspondence should be addressed:

Dr. Luke Coburn lukecoburn@gmail.com

Dr. Guillermo A. Gomez g.gomez@uq.edu.au

Running head: Modeling of epithelial cell clusters

Keywords: Epithelial cell mechanics, vertex model, junctional tension, contact inhibition of locomotion.

Abbreviations:
CIL, Contact inhibition of locomotion, MCS, Monte-Carlo step, TFM, Traction force microscopy.

**Abstract (200 words)**

We generated a computational approach to analyze the biomechanics of epithelial cell aggregates, either island or stripes or entire monolayers, that combines both vertex and contact-inhibition-of-locomotion models to include both cell-cell and cell-substrate adhesion. Examination of the distribution of cell protrusions (adhesion to the substrate) in the model predicted high order profiles of cell organization that agree with those previously seen experimentally. Cells acquired an asymmetric distribution of basal protrusions, traction forces and apical aspect ratios that decreased when moving from the edge to the island center. Our in silico analysis also showed that tension on cell-cell junctions and apical stress is not homogeneous across the island. Instead, these parameters are higher at the island center and scales up with island size, which we confirmed experimentally using laser ablation assays and immunofluorescence. Without formally being a 3-dimensional model, our approach has the minimal elements necessary to reproduce the distribution of cellular forces and mechanical crosstalk as well as distribution of principal stress in cells within epithelial cell aggregates. By making experimental testable predictions, our approach would benefit the mechanical analysis of epithelial tissues, especially when local changes in cell-cell and/or cell-substrate adhesion drive collective cell behavior.

**TOC (350 character)**

We generated a computational approach to analyze the biomechanics of epithelial cells based on their capacity to adhere to one another and to the substrate and exhibit contact inhibition of locomotion. This approach has the capacity to reproduce emergent properties of epithelial cells and make predictions for experimental validation.

**Introduction**

In epithelial tissues, the capacity of epithelial cells to alter their shape, move and exchange neighbors is profoundly influenced by the biochemical and mechanical properties of the tissue (Mammoto *et al.*, 2013; Lecuit and Yap, 2015; Mao and Baum, 2015). Adhesion, to either the substrate or to another cell, allows cells to probe and respond to the mechanical properties of their environment. At the sites of cell-cell junctions, adhesion receptors like cadherins, couple the contractile actomyosin apparatuses of epithelial cells together to generate junctional tension (Gomez *et al.*, 2011). Physical tension on junctions has been revealed by a variety of methods including laser ablation (Ratheesh *et al.*, 2012; Smutny *et al.*, 2015), optical tweezers (Bambardekar *et al.*, 2015), FRET tension sensors (Grashoff *et al.*, 2010; Borghi *et al.*, 2012; Conway *et al.*, 2013; Leerberg *et al.*, 2014) and immunofluorescence for protein epitopes that are revealed under tension (Yonemura *et al.*, 2010). In particular, FRET-based molecular tension sensors have been useful to show that both E-cadherin and vinculin molecules experience tension when localized at the epithelial cell junctions (Borghi *et al.*, 2012; Leerberg *et al.*, 2014). At the cell-substrate interface, integrin receptors interact with ligands in the extracellular matrix and exert forces on these adhesion sites, thus probing the mechanical properties of the substrate. This process allows the maturation and the recruitment of signaling and adaptor proteins to these adhesion sites (Grashoff *et al.*, 2010; Roca-Cusachs *et al.*, 2013).

Traction force microscopy (TFM) has been instrumental to measuring the direction and magnitude of forces that cells apply on their substrate (Saez *et al.*, 2010; Style *et al.*, 2014; Martiel *et al.*, 2015). When applied to clusters of epithelial cells and combined with Newton's law of force balance, this technique also allows the inferred measurement of "tugging" forces that occur on cell-cell junctions and the physical stress in the monolayer (Trepat *et al.*, 2009; Liu *et al.*, 2010; Maruthamuthu *et al.*, 2011; Tambe *et al.*, 2013; Ng *et al.*, 2014). In the case of a pair of cells, traction forces develop principally at the periphery of the cell cluster and are balanced with tugging forces exerted by cells at their cell-cell junctions (Liu *et al.*, 2010; Maruthamuthu *et al.*, 2011). Bigger cell clusters (>2-1000 cells) still show some similarities with a pair of cells with traction forces localized primarily at the periphery of the cluster (Trepat *et al.*, 2009; Mertz *et al.*, 2013; Ng *et al.*, 2014). However, under these circumstances, traction forces also develop in cells behind the border or leader cells that are located at the edges of the cluster, as these cells are able to form cryptic lamellipodia that extend underneath their neighbours (Trepat *et al.*, 2009). Video microscopy, on the other hand, has shown that higher velocities exhibited by leader cells at the edge of these multicellular aggregates correlate with an alignment in this direction of the principal stress vector in cells behind them, a phenomenon called plithotaxis, which has been implicated in collective cell migration (Zaritsky *et al.*, 2015). Finally, stress inference from TFM has also shown that stresses at the cell-cell junctions are higher in the island center and become smaller in the periphery where traction forces are higher (Trepat *et al.*, 2009; Mertz *et al.*, 2013; Ng *et al.*, 2014). Results from these experiments further revealed a mechanical crosstalk between both cell-cell and cell-substrate adhesion systems (Martinez-Rico *et al.*, 2010; Jasaitis *et al.*, 2012; Weber *et al.*, 2012; Mertz *et al.*, 2013).

Recently, the Prost lab developed a theoretical framework, incorporating adhesion between cells and their substrate, for the analysis of the physical behavior of epithelial sheets and how it defines different properties of the tissue in three dimensions (Hannezo *et al.*, 2014). In addition, particle-based simulation approaches have been used to model the dynamics of adhesive clusters that have been successful in predicting the pattern of forces developed by cell aggregates (Zimmermann *et al.*, 2014; Zimmermann *et al.*, 2016). However, these models lack important physiological features of cells, such as cell protrusions, cell-substrate adhesion and/or cell-cell junctions, limiting their ability to incorporate experimental data about these features. Finally, vertex and cellular Potts models have been extensively used to describe the physical state of epithelial cells. In particular, in vertex models cells are modeled as polygons and the position of the vertex is varied according to a probabilistic rule that depends on the cell elasticity, junctional contractility and cell-cell adhesion parameters (Farhadifar *et al.*, 2007; Fletcher *et al.*, 2014; Bi *et al.*, 2015; Misra *et al.*, 2016). On the other hand, in cellular Potts models a cell is made up of a given number of pixels that are connected and allowed to change the index (cell) that has been assigned to them so they go from belonging to one cell to belonging to another according to some probabilistic rule that frequently is very similar to the rules used to change vertex positions in vertex models (Kabla, 2012; Noppe *et al.*, 2015; Magno *et al.*, 2015; Albert and Schwarz, 2016). Together, vertex and cellular Potts models have been effective in describing how cell packing and total interfacial tension depends on basic features such as cell contractility and cell-cell adhesion (Farhadifar *et al.*, 2007; Noppe *et al.*, 2015). However, by themselves, these models of cell-cell junctions are not well suited for discrete systems with few cells, where adhesion to the substrate becomes more important as the size of the island becomes smaller.

Here we created a vertex version of our recently reported continuous model of confluent epithelial cells (Noppe *et al.*, 2015), where cells were also able to interact with the substrate and exhibit contact inhibition of locomotion (CIL, Coburn *et al.*, 2013). Using this model we analyzed the traction force, monolayer stress and junctional tension distribution of discrete epithelial systems (~10 to 300 cells) and made comparisons with experimental data. We found the model reproduced well the previous experimental observations on the distribution of traction forces and monolayer stress at cell-cell junctions as well as showing the presence of plithotaxis and mechanical crosstalk between both adhesion systems. It also predicted that junctional tension is not homogeneous along the island, but rather, scales up with island size, which we confirmed experimentally by using laser ablation. Thus, our model has the capacity to generate emergent properties of epithelial cells that can benefit the analysis of epithelial tissue mechanics.

**Results**

*Model of epithelial cells*

To build our model, we considered that the biomechanics of epithelial cells can be analyzed in terms of the behavior of a) their basal surface, which interacts with the substrate and forms protrusions; and b) the cell-cell interface, where cells adhere to one another and which exhibits contractile properties (Wu *et al.*, 2014). The model then takes into account that these two features are coupled mechanically through the body of the cell, which has some degree of intracellular stiffness (Fig.1a). This constitutes a first simplification to focus on two important features of epithelial cells. Although strictly speaking, the model does not aim to describe the three dimensional properties of cells, with the above simplifications, we propose a computational approach for the analysis of the biomechanics of epithelial cell aggregates based on the coupling of our previously reported algorithms for cell-cell adhesion and cell-substrate adhesion.

*Cell-cell adhesion and junctional contractility.*

A common way to model the apical cell-cell junctions and the apical surface of epithelial tissues is by the use of vertex models (Farhadifar *et al.*, 2007; Canela-Xandri *et al.*, 2011; Bi *et al.*, 2015). In this approximation, the tissue or cell aggregate surface is represented by connected polygons in the 2D plane where cell-cell interactions occur. Within this plane, each polygon corresponds to one cell, the edges between two polygons correspond to a cell-cell junction and vertices correspond to points where three or more cells meet. The energy $(E_i)$ for the $i^{th}$ cell

is then calculated as:

$$E_i = -J \sum_{j=1}^{n} l_j + K \sum_{j=1}^{n} l_j^2 + \lambda [a_i - a_o]^2 \quad (1)$$

where $a_i$ is its apical surface area, $l_j$ is the cell contact length between two cells, $n$ is the number of contacts that the $i^{th}$ cell make with its neighbors and $a_o$ is the preferred apical surface area for all cells. The parameters $J, K$ and $\lambda$ are the system parameters that weight the contribution of adhesion, junctional contractility and volume elasticity (at constant cell height), respectively. The first term in (1) is the cell-cell adhesion term and it becomes more negative as the perimeter elongates reflecting the capacity of cells to adhere to one another. The second term in (1) is related to junctional contractility that tends to reduce the contact length (and the cell's perimeter) and thus generates junctional tension. The third term in (1) relates to the cell's elasticity, where cells are allowed to have a variable shape, but their volume is kept constant (see below) by varying their apical area around a target area $a_o$. This last term finds a minimum when $a_i = a_o$, i.e. when cells acquire their preferred apical area.

The total energy of the system ($E_T$) for a given configuration of vertices is given by $E_T = \sum_{i=1}^{N} E_i$, where $N$ is the number of cells present in the aggregate. A typical simulation will start with cells configured into a square lattice and then, by following a Monte-Carlo algorithm, we update the vertex positions until a stable configuration is obtained. More specifically, in a single Monte-Carlo step (MCS), a vertex is randomly selected and one of the following process is performed: (i) the vertex is moved by a distance randomly selected from the range $[0, dr]$ in a random direction where $dr = 0.1$; or (ii) split into two vertices by defining a new vertex and hence generating a new bond connected to the chosen vertex (junction formation); or (ii) destroyed by selecting a bond and removing one of the vertices at its end points (junction removal). In each MCS, these three processes have equal probability of being selected at the same time that internal angles defined by two consecutive junctions in a cell are limited to the range [0, π]. After this change is made, the variation in the total energy of the system $\Delta E_T = E_T(after) - E_T(before)$ is calculated based on Eq. 1 and the update is made according to a Metropolis procedure: If $\Delta E_T < 0$, then the change has led to a reduction in energy and it will be accepted. If $\Delta E_T > 0$, the change can still be accepted with a probability $e^{-\Delta E_T/\chi}$, where $\chi$ is the noise parameter. To map the dynamics of the junctions at the apical surface onto the dynamics of the basal surface (see below) we assume that one time step corresponds to a Monte-Carlo cycle (number or vertex MCS attempts on random selected vertices). Simulations were performed with values of $a_o = 1$ and $\lambda = 0.5$ unless otherwise specified.

*Cell-substrate adhesion and cell motility.*

To introduce adhesion to the substrate and cell motility we modified our previous CIL algorithm (Coburn *et al.*, 2013). Briefly, in this model cells are allowed to adhere to their substrate, spread their basal area and extend protrusions in the direction that they migrate, similarly to real cells when they migrate into a free surface. If an asymmetry in the protrusions is present (i.e. a net force of traction on the cell exists), then the cell will move in the direction of the asymmetry (Caballero *et al.*, 2014; Wong *et al.*, 2014). We use this behavior to incorporate motility into our simulation (Coburn *et al.*, 2013). Then the contact inhibition of locomotion process (Roycroft and Mayor, 2016) is incorporated into the Monte-Carlo scheme as follows: if after a Monte Carlo Cycle the basal layer of two cells overlaps, cell protrusions are then retracted from the area of overlap in the radial direction towards their own cell center. This results in an alteration of the distribution of cell protrusions and a net change in the force of traction and cell orientation (see Fig. 1b).

In the simulations, time-averaged cellular protrusions are distributed uniformly around a cell. We represent this as a closed curve about a center point, $r_i^b(t)$ that is updated after a Monte Carlo Cycle. The initial basal perimeter $P$ is represented in polar form as (see also Fig. 1a):

$$P_i(\phi) = 0, \quad 0 < \phi < 2\pi \tag{2}$$

Cellular protrusions then relax to a time-averaged uniform distribution, $P_0(\phi) = A_1$ over the subsequent time steps. $\phi$ is represented as a discrete set of $m$ values with $\phi_j = 2\pi j/m$, $j = 1, \ldots, m$ and $m = 50$. To have an estimation of $A_1$, we performed different simulations of cell islands varying this parameter and measured the average basal to apical area ratios of cells. We then compared these values to those derived from experiments with the same number of cells in the island. We found that a value of $A_1 = 0.8$ generates an apical to basal area ratio in the simulations that fits with those observed experimentally.

In the numerical simulations protrusion contours are updated using discrete time steps where cells gradually remodel their protrusions around the target perimeter $P_0(\phi)$ according to

$$\frac{dP_i(\phi,t)}{dt} = -\gamma[P_i(\phi,t) - P_0(\phi)] + \xi(\phi,t) \tag{3}$$

where $\gamma$ determines the rate of regrowth. Random fluctuations are incorporated into the protrusion contour by adding an uncorrelated white noise function $\xi(\phi,t)$ with noise intensity $\rho$ where

$$\langle \xi(\phi,t)\xi(\phi,t+\tau)\rangle = \rho^2\delta(\tau) \tag{4}$$

Finally, to relate cell traction forces to their motility, we assume that cellular protrusions impart a net force on the cell in the direction of migration, whose force is proportional to the degree of asymmetry of the distribution of protrusions around the cell. For real cells, this assumption is valid within times scales where asymmetry in protrusions, cell velocity and the presence and direction of traction forces are correlated (Caballero *et al.*, 2014); but not on shorter time scales where there is a time delay between the occurrences of these phenomena (Notbohm *et al.*, 2016), which is related to the underlying mechanotransduction processes that occur at focal adhesions before cells exert forces on a new area of cell-substrate attachment (Wong *et al.*, 2014). With this assumption, we then define the total force that protrusions apply on the $i^{th}$ cell $\boldsymbol{F}_p^i(t)$, to be the integral of the protrusion lengths in all directions about the center of the cell:

$$\boldsymbol{F}_p^i(t) = h_o \int_0^{2\pi} P_i(\phi,t) u_\phi \, d\phi \tag{5}$$

where $h_o$ is a prefactor related to the capacity of cells to adhere to their substrate and cell motility (which also depend either on the presence of ligands and/or substrate mechanical properties), $u_\phi$ is the radial unit vector in the direction $\phi$ and $P_i(\phi,t)$ is the protrusion contour of the $i^{th}$ cell at time $t$.

*Coupling between-cell-cell adhesion and cell substrate adhesion and the contribution of intracellular cell stiffness to cell motility and apical cell interactions.*

Finally, to couple the apical and basal layers of the in silico cells, we approximate them as skewed prisms with parallel apical and basal surfaces with centroids $\boldsymbol{r}_i^a(t)$ and $\boldsymbol{r}_i^b(t)$, respectively (Fig 1a). These surfaces sit on top of a two-dimensional protrusion contour ($P_i(\phi,t)$) that determines the position of the basal centroid. Then, for a randomly moving cell, attachment to a neighbor limits its freedom and this can be represented as a damping of its motility. This damping also depends on the stiffness of the cell and how deformable it is, which ultimately depends on the properties of the cortical actin cytoskeleton. For this reason we include an extra term in the cell motility description to account for this effect. Although the cell cytoskeleton is an over-damped network of different biopolymers, we assume that over short time scales it has an intrinsic stiffness and behaves as an elastic spring with constant $s$, which determines how the force is transmitted through the cell interior. This is included in the above CIL model by introducing an additional spring term for the horizontal displacement between the $\boldsymbol{r}_i^a(t)$ and $\boldsymbol{r}_i^b(t)$ centroids

$$\frac{d\boldsymbol{r}_i^b(t)}{dt} = \boldsymbol{F}_p^i(t) - s\Delta \mathbf{r}(t) \quad (6)$$

where $\Delta \mathbf{r}(t) = \boldsymbol{r}_i^a(t) - \boldsymbol{r}_i^b(t)$. Thus, the second term in (6) acts against the force due to cell protrusions and limits the offset between the apical and basal surface of a cell. Using Eq. 6, position of the basal surface is updated following the first order Euler scheme:

$$\boldsymbol{r}_i^b(t+dt) = \boldsymbol{r}_i^b(t) + \frac{d\boldsymbol{r}_i^b(t)}{dt}\Delta t$$

where $\Delta t$ corresponds to a Monte Carlo Cycle (or one simulation time step).

Similarly, the intracellular cell stiffness is also incorporated into the apical layer (Eq. 1) by including a spring term ($cs|\Delta \mathbf{r}(t)|^2$) in the energy function:

$$E_i = -J\sum_{j=1}^{n} l_j + K\sum_{j=1}^{n} l_j^2 + \lambda[a_i - a_o]^2 + cs|\Delta \boldsymbol{r}(t)|^2 \quad (7)$$

where $c$ is an scaling factor. This term has a minimum when the distance between the horizontal displacement between the apical and protrusion center is zero ($|\Delta \mathbf{r}(t)| = 0$), which is the case for confluent epithelial cells layers analysed under periodic boundary conditions.

*Cell volume preservation.*

We performed two types of simulations depending on the boundary conditions: (i) periodic boundary conditions to model confluent monolayers, and (ii) non/semi-periodic boundary condition to model cell islands and stripes in which some boundary layer cells will "see" free space instead of another cell. In our simulations using periodic boundary conditions, we assume that the cell volume is conserved locally by changes in monolayer height as in (Farhadifar *et al.*, 2007). In contrast, for simulations with open boundaries (islands), the apical surface area is free to expand or contract. In order to conserve cell volume the height of the monolayer, which we assume to be constant for bulk cells, is varied. To better illustrate this point, in Fig. 1c we show an example of two islands with the same volume but with different apical surface area and height.

*Approach to Modeling Cell Stripes and Islands.*

As mentioned above we will consider the scenario in which some cells will not be completely surrounded by other cells. At the border between a cluster of cells and the free space one can expect a growth of cellular protrusions towards the free space (Poujade *et al.*, 2007; Trepat *et al.*, 2009). Thus, parameters defining the protrusion sizes, traction forces and the intrinsic cell stiffness can be adjusted using experimental data derived from small islands (< 10 cells). We varied the ratio of basal cell area to apical cell area and the absolute value of the average horizontal offset between apical and basal centroids to match parameters for protrusions between simulations and experiments. In addition, since islands of cells are no longer periodic (and their area is not fixed) the total apical area of stripes and islands can reduce or expand beyond the preferred apical area observed in confluent monolayers. For a better comparison between behavior of the cells at the different position of the island or stripe, averages of traction and tension are shown only for bulk cells (i.e. row cell number>1).

## Mechanics of confluent epithelial cell monolayers

*Hard and soft regimes within confluent epithelial cell monolayers.*

Junctional tension makes epithelial tissues more rigid. In the model, the presence of junctional tension is determined by the presence of positive interfacial tension (Magno et al., 2015), according to the following equation:

$$\frac{dE_i}{dl_i} > 0 \qquad (8)$$

Hard and soft regimes thus can be defined by the presence of positive or negative interfacial tension $\left(\frac{dE_i}{dl_i}\right)$, respectively. Although adhesion to the substrate is present in the model, for simplicity we can neglect its contribution in the analysis of confluent monolayers as under these conditions, protrusions symmetrically distribute around the cell-substrate interface and within a monolayer apical and basal centroids are effectively aligned; therefore $cs|\Delta r(t)|^2 = 0$. Supporting this assumption is the fact that the boundary between soft and hard regimes in confluent monolayers is not affected by varying the motility and cell stiffness parameters (Fig. 2d). Using this result and taking the derivative of Eq. 7, we then obtain

$$\frac{dE_i}{dl_i} = -J + 2Kl_i + 2\lambda[a_i - a_o]\frac{da_i}{dl_i} \qquad (9)$$

We then consider the case of epithelial monolayers formed by cells with regular polygonal shape (i.e. all $n$ sides equivalent $l_1 = l_2 = l_j = \cdots = l_n = l_i$), which can uniformly pack or tile a surface without leaving gaps. Although regular pentagons ($n = 5$) cannot tile uniformly a surface the equations below constitute an approximation for cells with irregular pentagonal shape. Thus, for regular polygons, it is possible to write the following relationships between polygon area $a_i$, perimeter $p_i$ and side (or cell-cell contact) length $l_i$ (Supplementary Figure 1, Staple *et al.*, 2010; Magno *et al.*, 2015):

$$\sum_{j=1}^{n} l_j = n l_i = p_i \qquad (10)$$

$$\sum_{j=1}^{n} l_j^2 = n l_j^2 = \frac{p_i^2}{n} \qquad (11)$$

and

$$a_i = \frac{1}{4n} \cot\left(\frac{\pi}{n}\right) p_i^2 \qquad (12)$$

Introducing the polygonal shape descriptor $\eta = \frac{n}{4} \cot\left(\frac{\pi}{n}\right)$, it is possible to then express

$$a_i = \frac{n}{4} \cot\left(\frac{\pi}{n}\right) l_i^2 = \eta l_i^2 \qquad (13)$$

Replacing these equalities into Eq. 9, we then obtain

$$\frac{dE_i}{dl_i} = -J + 2K l_i + 4\lambda a_0 \left(\frac{a_i}{a_o} - 1\right) \eta l_i \qquad (14)$$

For confluent epithelial monolayers Eq. 14 could be further simplified by considering that cells cover the entire underlying surface and each cell acquire on average an area $\bar{a}$ given by:

$$\bar{a} = \frac{A_T}{N} \qquad (15)$$

where $A_T$ is the total area that the cells cover and $N$ is the number of cells in the system. Note also that $a_0$ is the preferred area of cells and thus is possible to define a factor $\zeta$ that relates to monolayer confluence (or cell packing density).

$$\zeta = \frac{\bar{a}}{a_0} \qquad (16)$$

When $\zeta > 1$, cells cover the surface but they do it by extending their area above the preferred area (this scenario correspond to the case where the number of cells is suboptimal to cover the surface). On the other case, when $\zeta < 1$, cells are densely packed and their average area is below their preferred area.

With the above last consideration, it is then possible to rewrite Eq. 14 as:

$$\frac{dE_i}{dl_i} = -J + 2K l_i + 4\lambda a_0 (\zeta - 1) \eta l_i > 0 \qquad (17)$$

that corresponds to the zero order Taylor approximation at $l = \sqrt{\frac{a_0}{\eta}}$. In Eq. 17, the equilibrium cell-cell contact length, $l*$, a junction that is not under tension or compression, is given by:

$$\frac{dE_i}{dl_i} = -J + 2Kl_i^* + 4\lambda a_0(\zeta - 1)\eta l_i^* = 0 \quad (18)$$

While one of the solutions to Eq. 18 corresponds to the simple case where cells are optimally packed, i.e. $\bar{a} = a_0$ and $\zeta = 1$ so then $l_i^* = l_0 = \frac{J}{2K}$; its more general solution is:

$$l_i^* = \frac{l_0}{1 + \frac{2\lambda a_0 \eta(\zeta - 1)}{K}} \quad (19)$$

For physically meaningful solutions we require $l_i^* > 0$. Thus,

$$\zeta > 1 - \frac{K}{2\lambda a_0 \eta} > 0 \quad (20)$$

Now, It is also possible to solve Eq. 17 to determine when cells within a monolayer with a given density and polygonal arrangement, which have junctions of length $l_i$, will have a positive value of junctional/interfacial tension ($\frac{dE_i}{dl_i} > 0$). This is given by meeting the following condition:

$$l_i > l_i^* = \frac{l_0}{1 + \frac{2\lambda a_0 \eta(\zeta - 1)}{K}} \quad (21)$$

If we consider a confluent monolayer covered uniformly by polygons of the same shape, Eq. 21 can be rewritten as:

$$l_i = \sqrt{\frac{\bar{a}}{\eta}} = \sqrt{\frac{a_0 \zeta}{\eta}} > \frac{l_0}{1 + \frac{2\lambda a_0 \eta(\zeta - 1)}{K}} \quad (22)$$

Replacing $l_0 = \frac{J}{2K}$ and rearraging to solve for K then for the hard regime we obtain:

$$K > \sqrt{\frac{\eta}{4a_0\zeta}} J - 2\lambda a_0 \eta(\zeta - 1) \quad (23)$$

and the expression

$$K^0 = \sqrt{\frac{\eta}{4a_0\zeta}} J - 2\lambda a_0 \eta(\zeta - 1) \quad (24)$$

defines the line in the phase diagram that delimits the presence of cells with positive or negative interfacial/junctional tension.

*Numerical simulations of confluent epithelial cell monolayers.*

To validate the model, we perform simulations of confluent monolayers to compare the model's behavior to the above theoretical predictions (Fig. 2). First, to characterize the amount of junctional/interfacial tension, we analyze the net force that is exerted on cell vertices when individual cell-cell junctions are removed (Fig 2a). Using this definition, negative values of force correspond to junctions under compression whereas positive values denote junctions under tension. For each junction and configuration, we calculate an ensemble average of junctional tension, by first removing a cell junction and then calculating the change in the energy of the system $dE_T$, after dragging apart its vertices by an amount $dl$. Thus the tension on each junction is calculated as $\langle T \rangle = \langle -\frac{dE_T}{dl} \rangle$ (Fig. 2a), for which the distance $dl$ is reduced until the values obtained for the tension converge. This approach is comparable to the experimental situation where a cell-cell junction is cut by laser ablation (Gomez *et al.*, 2015).

Simulations were performed varying the parameters that control cell-cell adhesion energy $(J)$ and junctional contractility $(K)$. In agreement with previous simulations using vertex and cellular Potts models (Farhadifar *et al.*, 2007; Noppe *et al.*, 2015), we found that for high contractility/adhesion $(K/J)$ ratios, cells acquire regular hexagonal order, whereas when the adhesion term is more prominent (low $K/J$ values) the regular packing of cells is lost (Fig. 2b). We then calculated the average junctional tension for the contacts in the lattice as a function of the adhesion $(J)$ and contractility $(K)$ parameters to create a phase diagram (Fig. 2c) and compared it with our theoretical predictions (Eq. 24). We found that regions where the packing is more regular correspond to overall high junctional tension (hard regime), whereas irregular packing with less ordered polygonal shapes is observed in the systems having lower junctional tension (soft regime, Fig 2c). We also noticed that there is a boundary between regions with positive and negative junctional tension, which is in excellent agreement with our theoretical description and the predictions using Eq. 24 (red dotted line in the phase diagram).

To further characterize the system and to define whether a phase transition occurs from the soft to the hard regime when contact contractility increases, we performed simulations with a constant cell-cell adhesion parameter $(J = 0.375)$ while increasing contractility $(K)$ systematically (see Fig. 2d). We observed that there is a transition in the amount of junctional tension at contractility values $K \sim 0.3$ that compares well with the analytical prediction of using Eq. 24 (Fig. 2c). This suggests that, under these conditions, the effect of increasing contractility not only rigidifies the entire system but also collectively affects epithelial cell organization.

We then attempted to elucidate the role of cell-substrate interactions and cell propulsion in the onset of hard and soft regimes in confluent monolayers. For this we performed the simulations presented in Fig. 2d, varying the cell substrate interaction term $h_0$ that defines the speed at which cells can move in the absence of cell-cell adhesion (Coburn *et al.*, 2013). We found that introducing motility to cells does not alter the qualitative behavior of the model in simulations of confluent cell monolayers. A similar result was obtained when the cell stiffness parameter ($s$) was modified. Altogether, the results of the model suggested that within confluent monolayers cell motility neither contributes to increasing the forces on cell-cell junctions nor the mechanics of cell-cell junctions. This response results from the fact that under these conditions no net cell asymmetry in the basal layer is favored and therefore the term $cs|\Delta r(t)|^2$ becomes negligible.

To further evaluate the performance of our modeling approach, we performed experiments (see section Experimental Procedures for more details) where we analyzed the morphology of confluent cell monolayers treated with the myosin II inhibitor, blebbistatin (Blebbi) or DMSO vehicle (control, Fig. 3a). We then compared our empirical results to four *in silico* cases: Case 1: $J = 0.375, K = 0.2$ (soft regime); Case 2: $J = 0.375, K = 0.45$ (hard regime with high adhesion); Case 3: $J = 0.075, K = 0.5$ (hard regime with low adhesion energy, see also Fig 2c) and Case 4: $J = 0.075, K = 0.2$ (hard regime with low adhesion and contractility). We recorded the cell shape distribution and polygon number of cells in simulations and experiments and compared the ratio of average area in a polygon class to average area, $\langle a_n \rangle / \langle a \rangle$ vs polygon number as it was described before (Farhadifar *et al.*, 2007; Canela-Xandri *et al.*, 2011, Fig 3b). We found that in control cell monolayers the rate of change of $\frac{\langle a_n \rangle}{\langle a \rangle}$ with the polygon number is similar to cases where cells exhibit high junctional tension (case 2 and, to a limited extent, case 3, Fig 3b), in agreement with the fact that under normal circumstances cell junctions are under tension (Wu *et al.*, 2014; Gomez *et al.*, 2015). Moreover, we found that when myosin II is inhibited, the rate of change of $\frac{\langle a_n \rangle}{\langle a \rangle}$ with the polygon number behaves similarly to control cells (case 2), suggesting that in addition to an inhibition in junctional contractility, cell-cell adhesion energy could be also compromised under these experimental conditions. This notion agrees with earlier evidence that myosin activity and junctional tension are required for the stability and accumulation of E-cadherin adhesion molecules on cell-cell junctions (Shewan *et al.*, 2005; Rauzi *et al.*, 2010; Smutny *et al.*, 2010) and with results of numerical results in which we lower adhesion as well as contractility (Case 4, Fig 3b) and observe a similar rate of change of $\frac{\langle a_n \rangle}{\langle a \rangle}$ with the polygon number to the case of cells with higher adhesion and contractility (Case 2). Thus, our modeling approach correlates well with the behavior observed in confluent monolayers of epithelial cells.

### *Mechanics of epithelial cell Islands and stripes*

*Force distribution in epithelial cell islands and stripes.*

To quantify the stress distribution in small epithelial cell aggregates, we modeled group of cells forming stripes and islands (see Fig. 4a) and analyzed the steady state distribution of monolayer stress and traction forces from the edge to up 10 cell diameters (rows) inside the cluster. The total traction that epithelial cell islands or stripes apply on their substrate is calculated as a function of the distance from the edge ($r$) as the average projection of $\boldsymbol{F}_p^i$ (Eq. 5) in the horizontal or radial directions for stripes and islands, respectively (see also Fig. 4b):

$$Tr(r) = \langle \boldsymbol{F}_{p,r} \rangle \qquad (25)$$

The presence of traction forces generates physical stress across the cell monolayer that is transmitted through the cytoskeleton and cell-cell junctions (Trepat *et al.*, 2009; Liu *et al.*, 2010; Maruthamuthu *et al.*, 2011; Ng *et al.*, 2014). At a specific position $r$ from the edge, these traction forces are balanced by the local stresses in the cell monolayer (Trepat *et al.*, 2009). Therefore, at a specific distance *r* from the edge the sum of traction forces is balanced by the local stress in the monolayer at that position, thereby allowing us to calculate this monolayer stress in the model as a function of the distance from the cell edge as

$$\sigma(r) = \sum_{r=0}^{r} Tr(r) \qquad (26)$$

We found that for both cell stripes and islands, traction forces are higher for cells at the edge and lower for cells at the center. Moreover, we found that there are still significant (although smaller) traction forces for cells located in the third and fourth row behind the edge (Fig. 4c, see also Fig 6a,i and b,i), suggesting that these cells still have the capacity to pull the island in the outward direction. When the monolayer stress profiles were analyzed, we found that for islands and stripes, the stress is higher at the center but lower in the periphery (Fig. 4c). Overall the pattern observed for traction forces and monolayer stress for cell islands correlated well with previous results obtained experimentally with similar cell cultures (Trepat *et al.*, 2009;

Maruthamuthu *et al.*, 2011; Ng *et al.*, 2014).

Based on these results we then investigated how those cells away from the island edge experience net traction forces. Cells at the edge have no neighbors outside the island and their protrusions can extend more into the free space. Such asymmetry could be propagated, to some extent, inside the island, thus allowing cells in this location to generate traction forces. To analyze whether this is the case, we examined the degree of asymmetry between apical and basal cell areas ($|\Delta \mathbf{r}|$) as a function of the distance from the edge, as a measure of the formation of cryptic lamellipodium in the model (Fig. 4d). Similar to the traction force data, we found that the cells in the model exhibit a notable degree of asymmetry when located in proximity to the island and stripe edge. This degree of asymmetry is more pronounced in the direction orthogonal to the island edge than in the direction parallel to it (Supplementary Figure 2a) and decays with the cell position from the edge (Fig. 4d, Supplementary Figure 2a), similarly to what is observed in TFM of cell islands and during collective epithelial cell migration (Trepat *et al.*, 2009; Das *et al.*, 2015). We also considered the morphology of the apical region of cells and how the apical area orientates with respect to the radial direction of the island. The results in Fig. 4e show that cells preferentially elongate their apical area in the direction orthogonal to the island edge, meaning that its longer axis is oriented along the island's radial direction (average $\cos(\theta) \sim 0.4$ compared to average $\cos(\theta)=0$ for a randomly oriented cells). Moreover, this degree of orientation penetrates several cell diameters within the island from its edge (Supplementary Movie 1). This resembles the phenomenon of plithotaxis during collective epithelial migration where the direction of cell's principal stress is parallel to the direction of cell's velocity (Zaritsky *et al.*, 2015), which in our case is defined by the CIL process. Of note, we also noticed that in the model the formation of asymmetries in cell protrusions and their extension inside the island or stripe are positively regulated by the motility of cells and negatively regulated by the cell stiffness (Supplementary Figure 2b and c). Altogether, these results agree with experimental analysis of lamellipodium formation during epithelial cell migration (Abreu-Blanco *et al.*, 2012; Anon *et al.*, 2012; Das *et al.*, 2015); the observed of cryptic lamellipodia underneath cells located towards the edge of the island (Farooqui and Fenteany, 2005; Trepat *et al.*, 2009); and the alignment of the apical area of cells in the velocity direction as it is observed during plithotaxis (Zaritsky *et al.*, 2015).

As the model predicts the pattern of stresses that are similar to those observed experimentally, we then investigated what property in the model accounts for the generation of this monolayer stress. We rationalize that the stress in the apical region of cells could be generated by the resistance of cells to deform their apical area or by an increase in the amount of tension at cell-cell junctions. To analyze this last possibility, we performed numerical calculations of junctional tension as described in Fig 2a and plotted the results versus the distance of the particular junction from the island or stripe edge (Fig 4f). We found that under these conditions, junctional tension is lower in the peripheries of island and stripes and higher in the centers, thus having a similar profile to the average stress in the monolayer and suggesting that cell junctions contribute significantly to monolayer stress in the model. If this would be the case in real cells, then our simulations predict that junctional contractility should exhibit the same pattern when analyzed in epithelial cell islands, i.e. be lower in the periphery and higher in the center of the island. To test this, we analyzed the phospho-myosin regulatory light chain (pMRLC) content at cell-cell junctions in epithelial cell islands as a proxy for junctional contractility. In agreement with the model predictions, we found that pMRLC junctional content was lower at the island periphery and higher in the center (Fig 4g-i).

*Effect of island size on epithelial mechanics.*

Having found that the monolayer stress and junctional tension are higher at the island center, we used the model to explore whether or not this also depends on island size and compared this result with experimental data. For this, we performed simulations of islands of different sizes and measured the stress and junctional tension at the center. We found that these parameters quickly scale up in islands of radius from 2 to 6 cell layers and then reach a plateau for bigger cell aggregates (see Fig. 5a,b). We then performed laser ablation experiments on cell junctions located at the center of epithelial cell islands of different size to test the model's prediction. For this, epithelial cell islands were grown to different sizes from single cells and initial recoil after laser ablation was measured on a single junction per island as described previously (Gomez *et al.*, 2015). In experiments using this assay, the measured amount of tension on junctions at the center of the islands increased with the size of the island, a result that agreed with the predictions of our model (Fig. 5c). Thus our in silico and experimental results suggest that in cells the amount of junctional tension is a collective emergent property of the system.

*Mechanical crosstalk between cell-substrate and cell-cell adhesion sites.*

We then investigated whether or not junctional contractility and cell motility influence the patterns of traction force and junctional tension exhibited by epithelial cell islands, thus testing if the model exhibits mechanical crosstalk between the cell-cell and cell-substrate adhesion systems.

We first performed a set of simulations varying the cell motility parameter and analyzed the distributions of traction force, monolayer stress, offset of apical and basal areas, and junctional tension (Fig. 6a). As expected, we found that increasing cell motility leads to an increase in the amount of traction force and increases the offset between the apical and basal areas in the cells at the periphery of the island (Fig. 6a i and ii). More surprisingly, we found that the model exhibits some degree of mechanical crosstalk as the amount of junctional tension also increased when cell motility increased (Fig 6a iii). This suggested that in discrete systems of epithelial cell islands, cell motility contributes to the amount of forces that operate at cell-cell junctions.

We then also asked whether increasing junctional contractility in the model led to changes in traction forces and monolayer stress. We found that increasing junctional contractility leads to an increase in the amount of traction forces in the island periphery as well as an overall increase in the monolayer stress and junctional tension (Fig. 6b i and iii). This occurs together with an increase in the offset between apical and basal areas (cryptic lamellipodia index, Fig. 6b ii).

These results show that a simple model that minimally integrates cell-cell adhesion and cell motility together with CIL produces a strong crosstalk between the adhesion systems and show how this interaction leads to different emergent properties of epithelial cells that have been observed experimentally. Overall, these observations correlate well with the fact that cell-ECM traction force modulates junctional tension (Liu *et al.*, 2010; Maruthamuthu *et al.*, 2011) and that pulling forces on cadherin-junctions lead to an increase in cellular traction forces (Weber *et al.*, 2012; Mertz *et al.*, 2013).

**Discussion**

We have developed a quasi-three dimensional model of epithelial cells that includes their capacity to adhere to each other and to the substrate and exhibit contact inhibition of locomotion. In addition, cells in the model present intracellular stiffness and constant volume. Without being formally a three dimensional model, our approach has the analytic advantage that allows the direct extraction and comparison of cellular properties that can be measured experimentally. Parameters of the model like tension, cell motility, cell packing and topology can also be introduced into the model from experimental data. In particular, this aligns very well with many cases where analysis of mechanical properties of epithelial cells are made assuming that epithelial cells form a flat monolayer and this structure does not escape to the third dimension (e.g bends). Thus, although limited, this is a normal assumption that is made in experiments and our model naturally captures it.

Our approach allows straightforward comparison between simulations and experimental data as many of the biophysical methods that are suited to assess the biomechanical state of epithelial cells in monolayers are based on approaches that test how cells interact with their substrate and/or approaches focused on the apical layer of cells (Polacheck and Chen, 2016). Examples of these methods include the use of TFM with monolayer stress inference calculations (Nier *et al.*, 2016), and junctional tension that can be measured by laser ablation (Caldwell *et al.*, 2014) or optical tweezers (Bambardekar *et al.*, 2015). The model also allows comparison of the topology of cell arrangements within the monolayer between simulation and experiment, which allow mechanical inference of the state of cells even without directly measuring tension and/or stress, thus complementing other methods of stress inference (Chiou *et al.*, 2012; Sugimura and Ishihara, 2013). Accordingly, the model can be used not only to test predictions but it can be also fed with experimental results obtained from these experiments. This contrasts with other models that only take into account cell-cell interactions for example (Farhadifar *et al.*, 2007; Kabla, 2012), which work relatively well for local analysis of cells in relatively large tissues with negligible cryptic lamelipodia thus matching the conditions of the confluent monolayer cases analyzed in this work. Our model also contrasts with recent models that represent cells as particles (Zimmermann *et al.*, 2016), which are computationally robust and can make good predictions but lack the capacity to use experimental snapshots of cell topology to infer the mechanical properties of the tissue. Finally, although our modeling approach does not necessarily represent cells in three dimensions, it is a practical and straightforward application for most of the available experimental data that is not necessarily 3D. Of note, modeling in three dimensions requires extra measurements of parameters, which are not always possible to obtain directly or might be very difficult to measure, as are the variations in adhesion or tension in the XZ plane of the cell-cell interface (Wu *et al.*, 2014).

Using our approach we then developed a theoretical analysis complemented by numerical simulations on the presence of soft and hard regimes in confluent monolayers that correspond to cells exhibiting positive or negative values of interfacial/junctional tension (Farhadifar *et al.*, 2007; Noppe *et al.*, 2015; Magno *et al.*, 2015). Results from the simulations showed perfect agreements with our theoretical predictions. It should be noted that the soft/hard transitions described here are different from the jamming type of transition described recently by the Manning group (Bi et al, 2016) and observed experimentally (Park et al., 2015), where the energetics associated with T1 transitions are described as a change in polygonal arrangement, from hexagonal to pentagonal, within the vertex model. Thus, analyzing how adhesion to the substrate as well as CIL, in expanding cell islands or during collective migration, affects this type of transition constitutes new areas that deserve further research.

Then we applied our model to analyze the mechanics of epithelial cell stripes and islands. We have shown that for discrete systems, the presence of a free boundary can polarize cell protrusions at the edge of the island and this effect is propagated into the tissue to distances of several cell diameters. Due to this, islands and stripes develop patterns of traction forces, monolayer stress and junctional tension that vary from the edge to the center of the multicellular aggregate. We further found that these patterns agreed very well with the experimental observations and with a very recent report using a particle-based simulation model (Zimmermann *et al.*, 2016). Interestingly, our model allowed us to test the prediction that junctional tension at the center of islands increases with island size, which we confirmed experimentally using laser ablation. This can be explained, based on the mechanics of epithelial cell islands as the amount of radial outward traction that is generated in an island is proportional to the number of cells at the edge, i.e. to the island perimeter, which scales linearly with the island radius and as a square root of the area or the number of cells (at least for small islands). As at the same time, the amount of stress decreases away from the edge of the island due to internal damping, this explains why it would not grow any further when the size of the island becomes very large. This is in agreement with what was also shown before about the dependence of traction forces with the size of the epithelial cell aggregate (Mertz *et al.*, 2012). Thus, our modeling approach is able to unify a previously used continuous approaches with a more discrete model in which individual cell properties are now explicitly incorporated into it.

Similarly, our modeling approach showed that it has also the minimal properties that allow mechanical crosstalk between cell-cell and cell-substrate adhesion systems. Indeed, our simulations showed how the presence of developing traction forces are sufficient to increase junctional tension acting upon cells behind the edge of the island. Similarly, they showed how the presence of junctional contractility can also modulate the amount of traction forces that cells exert on their substrate which has been also observed experimentally (Jasaitis *et al.*, 2012; Weber *et al.*, 2012; Mertz *et al.*, 2013).

Our results on islands and stripes agree with the previous analytical description of traction forces exerted by contractile cell layers made by (Edwards and Schwarz, 2011; Banerjee and Marchetti, 2012). In their description, they found that higher traction forces are developed at the rim of the islands/stripes and this result naturally from the solution of a finite-sized contractile layer coupled to an elastic substrate. In our model, these two properties are held by the CIL component that allows the island/stripe to exert forces that try to increase the size of the island (similar to pillars in the description by Edwards and Schwarz, 2011) and the contractile cells with volume conservation and contractile junctions that tends to shrink the island or stripe.

Also, our modeling framework recapitulates some of the plithotaxis properties exhibited by epithelial cells (Zaritsky *et al.*, 2015). It has been shown that there is a correlation between the directions of migration and of cellular principal stress, which have been recently confirmed experimentally. In our model, CIL on cells on the edge of the island allow those cells to polarize and maximize their velocity vector in the outward direction. Close inspection of the snapshot of the simulation also showed that this lead to an increase aspect ratio of the cell in the radial direction that is propagated several cell diameters inside the island (Fig. 4e). This aspect ratio in the apical area of cells corresponds to the principal stress direction (as the basal layer does not contribute to it as it does not have cell-cell interaction). We also observe a similar effect in simulations of expanding cell islands, i.e. when simulations were initialized with cells having an average cell area smaller than the preferred area, and where we observed finger formation on some leader cells and the alignment of major stress towards this direction for cells located several cell diameters behind those leader cells (Supplementary Movie 1). Our model however does not make a distinction between cell motility and traction and indeed both vectors are aligned. Although this could be negligible in long time scales, recent analysis shows that on very short times scales this might not be well correlated (Notbohm *et al.*, 2016).

Our model also allowed us to investigate the propagation of forces within cell aggregates. In particular, we analyzed the effect of introducing four hyper contractile cells at the center of the epithelial cell island and analyzed how far force can be propagated. In agreement with previous experimental results (Ng *et al.*, 2014), we did not observe a significant degree of force propagation within the island, which was limited to one to two cell diameters (Supplementary Figure 2d). This again highlights the mechanics of epithelial islands as an emergent property of the system where the presence of adhesion to the substrate limits the capacity of the tissue to exhibit force propagation within the system as is discussed in more detail in (Ng *et al.*, 2014)

Although our modeling approach adapts very well for the analysis of epithelial cell organization in relatively flat tissues (2D-like systems), it is not suitable for the analysis of monolayer morphology in 3 dimensions. In particular, Vertex models have been extended to 3 dimensions to analyze the process of epithelial cell-cell rearrangements that occurs during cell extrusion and cyst formation (Bielmeier *et al.*, 2016) and during ventral furrow invagination (Polyakov *et al.*, 2014). Introducing CIL into these models would allow a more comprehensive view of the interplay between adhesion systems during these processes that involves force propagation, cell rearrangements and collective migration, whose theoretical analysis have already started to highlight the role of mechanics during these three dimensional morphological changes in tissues (Hannezo *et al.*, 2014).

Finally, so far our model of epithelial cells recapitulates only passive mechanical properties of cells and points out a key role of contact inhibition of locomotion as a spatial clue that leads to the polarization of cell mechanics within epithelial cell aggregates. Such polarization seems to be a general characteristic of cells and has also been introduced to reproduce experimental observations in the context of wound healing and collective migration (Banerjee *et al.*, 2015; Notbohm *et al.*, 2016). Although described as a passive element in our model, it has become increasingly clear that CIL is an active process and very recently, it has been found that the switching of cadherin cell-cell receptors plays a key role in CIL (Scarpa *et al.*, 2015). In particular, the work by Scarpa et. al. shows that the presence of E-cadherin contacts suppress CIL and its loss (or switch to N-cadherin) activates Rac1 signaling thus allowing cells to exhibit CIL. In the context of cell islands, this could correlate with different levels of Rac activity at the cell-substrate interface that is higher in cells in the periphery of the island and lower in those cells located in the center. This could lead to a pattern of protrusive activity and CIL that is consistent with our numerical simulations. Moreover, our data show that pMRLC content is higher in the island center but lower in the island periphery, but how cells control this at the molecular level is less clear, and perhaps reflects the capacity of E-cadherin based junctions to support mechanotransduction (Gomez *et al.*, 2011). Therefore, now becomes important to characterize the mechanotransduction mechanisms that cells use to regulate both the amount of force that acts on these adhesion systems and CIL, that results in the pattern of traction forces and junctional tension observed both theoretically and experimentally. This will help us to understand how cells resist the increasing stresses (and thus preserve tissue integrity) that occur in response to local changes in cell mechanics and drive collective cell behavior and morphological transitions that occur, for example during cell extrusion, wound healing and cell migration.

**Material and Methods**

*Immunofluorescence, microscopy and analysis of cell morphology.* Cells were fixed with 4% paraformaldehyde (PFA) in cytoskeletal stabilization buffer (Leerberg *et al.*, 2014). Immuno-staining was performed using rabbit anti phospho-(Ser19)-MRLC ab (Cat#36755, Cell Signaling) , Rat anti E-cadherin ab (ECCD-2, cat#13-1900, Invitrogen), mouse anti-ZO-1 (cat# 33-9100, Invitrogen) and Alexa conjugated secondary antibodies ( Invitrogen) as appropriate. Coverslips were mounted in Prolong Gold with DAPI (Cat#8961, Cell signaling). Confocal images were acquired on LSM 710 laser scanning microscopes (63x, 1.4NA Plan Apo objective) driven by Zen software (ZEN 2009, Zeiss). Images of control and blebbistatin (US1203390, Merck; 100 µM, 2 h) treated cell monolayers stained against ZO1 were used to obtained histograms of cell morphology using the packing analyzer 2.0 software (Aigouy *et al.*, 2010).

*Quantitation of pMRLC fluorescence intensity in epithelial cell islands.* For the quantitation of the radial distribution of pMRLC intensity in epithelial cell islands we used the radial profile extended plugin for image J by Philippe Carl (Laboratoire de Biophotonique et Pharmacologie, CNRS). Basically, a circle was drawn to enclose the edges of the island and a sector with identical island radius (as overall islands are not fully symmetrical) was selected to measure the average intensity of pMRLC content. The obtained profiles were then normalized to the average intensity at the edge of the island and the profiles then averaged. Radial distances were rescaled to 0 (island edge) and 1 (island center) to normalize profiles across different islands. Data corresponds to the average of 10 islands and values are mean ± SEM.

*Laser ablation experiments.* Cells were grown to confluence on glass-bottom dishes and cell media was replaced with Hank's Balanced Salt Solution (HBSS, Sigma) containing 5% FBS, 10mM HEPES pH 7.4 and 5mM $CaCl_2$ prior to imaging. The use of the laser ablation technique to assess junctional tension has been described previously (Gomez *et al.*, 2015). To assess junctional tension in steady-state monolayers (Figure 5), cells stably expressing Ecad shRNA/Ecad-GFP were used to identify the apical region of cell-cell contacts. All ablation experiments were carried out at 37°C on a Zeiss LSM510 system (40x, 1.3NA Oil Plan Neofluar objective) using 17% transmission of the 790nm laser on a 1 µm x1 µm area on the apical junctions of cells. Time lapse imaging of a 75 x 75 µm region was taken at 1.6 sec intervals before (3 frames) and after (42 frames) ablation.

Data was analyzed in ImageJ, using the MTrackJ plugin to track and measure the strain or deformation ε(t) of the cell-cell junction as a function of time after ablation. Since in the time scales of our experiments junctional strain exhibit a single exponential growth with a defined plateau, this was then modeled as a Kelvin-Voigt fiber (Fernandez-Gonzalez et al., 2009; Michael et al., 2016) by fitting it to the following equation

$$\varepsilon(t) = L(t) - L(0) = \frac{F_0}{E} \cdot \left(1 - e^{-\left[\left(\frac{E}{\mu}\right)*t\right]}\right)$$

where L is the length of the ablated junction measured as the distances between the vertices that define it, $F_0$ is the tensile force present at the junction before ablation, E is the elasticity of the junction and µ is the viscosity coefficient related to the viscous drag of the media. As fitting parameters for the above equation we introduced : $initial\ recoil = \frac{d\varepsilon(0)}{dt} = \frac{F_0}{\mu}$ and $k = \frac{E}{\mu}$

This model was used to calculate the initial recoil (the rate of recoil at t=0) for each junction ablated.

**Acknowledgements**
We thank all our lab colleagues for their support and advice. This work was supported by grants from the National Health and Medical Research Council of Australia (1067405 to AY, GAG and ZN; 1037320 to AY). ASY is a Research Fellow of the NHMRC (1044041). LC was funded under the Programme for Research in Third Level Institutions (PRTLI) Cycle 5 and co-funded by the European Regional Development Fund (ERDF). ZN is supported by an Australian Research Council Future Fellowship. Optical imaging was performed at the ACRF/IMB Cancer Biology Imaging Facility, established with the generous support of the Australian Cancer Research Foundation.

**Figure Legends**

**Figure 1. Model of epithelial cells.** a) In the model, cells are represented as skewed prisms sitting on top of a zero-volume contour (P(θ,t)) representing the cell-substrate interface (basal protrusions). The horizontal distance between the apical ($r_a$(t)) and basal ($r_b$(t)), is measured. b) The CIL interaction results in a retraction of the overlapping segments in the radial direction for each cell thus breaking the symmetry of the protrusion contour for each cell. As a result, both cells gain a net traction in the direction of the asymmetry. c) Epithelial cells are free to expand and contract. However, varying the height of the apical surface conserves the total volume of the monolayer.

**Figure 2. Mechanics of confluent monolayers**. a) Schematic of tension calculation. Once simulations reach its steady state, a cell-cell junction is selected and removed and the total energy of the system is calculated. The junction is then extended by a distance $dl$ and the resulting change in energy measured. Tension is then calculated as $T = -\frac{dE_T}{dl}$. b) Freeze frame of borderless monolayers for three values of contractility $K = 0.2, 0.3$ and $0.4$. c) Phase diagram of tension varying adhesion ($J$) and contractility ($K$). The highest junctional tension corresponds to higher values of the contractility and lower values of adhesion terms. The red dotted lines correspond to the theoretical limit between hard and soft regimes according to Eq. 24. The green line correspond to constant adhesion $J = 0.375$. d) Simulations varying contractility (K values) at constant adhesion ($J = 0.375$, green line in c). Cell substrate adhesion ($h_0$) and stiffness ($s$) parameters were also varied as indicated in the inset table.

**Figure 3. Analysis of cell morphology in confluent monolayers.** a) Images of control and blebbistatin treated (100µm, 2hs) confluent MCF-7 cell monolayers stained against the tight junction protein ZO-1. b) Cell shape distribution in the experiments was measured as described in material and methods and compared with those obtained in simulations (see also Fig 3c) by plotting the mean apical area in a polygon class over mean apical area ($\frac{<a_n>}{<a>}$) vs polygon number. Case I: $J = 0.375, K = 0.2$, Case II: $J = 0.375, K = 0.45$, Case III: $J = 0.075, K = 0.5$, and Case IV $J = 0.075$ and $K = 0.2$.

**Figure 4. Mechanics of epithelial cell islands and stripes.** a) Cell stripes and islands in simulation. b) The magnitude of horizontal/radial traction $|F_{p,r}|$ is calculated

by projecting the cell traction vector $F_{p,r}$ onto the horizontal/radial direction $r$. c) Traction and monolayer stress in the radial direction across the island and stripe. d) Plots of horizontal displacement (offset) between apical and basal centroids $|\Delta r(t)|$ across islands and stripes. e) Plot of the $|cos(\theta)|$, the angle between cell apical area long axis and the island radial direction across an epithelial cell island. f) Plots of junctional tension across Island. g-h) Immunofluorescence of MCF-7 epithelial cells islands stained against pSer19MRLC (green), E-cadherin (red) and Nuclei (DAPI, cyan) (g) and a magnification of the region indicated by the white square in f is shown in (h). i) Quantitation of pSer19MRLC (pMRLC) radial content as a function of the distance from the island edge.

**Figure 5. Effect of island size on its biomechanics.** Basal stress (a) and junctional tension (b) at island centre v island size (in rows) calculated from simulations. c) Initial recoil/junction length measured fat the center of MCF-7 cell islands of different size.

**Figure 6. Mechanical crosstalk of adhesion systems in the model.** Plots of i) traction, basal stress, ii) apical/basal offset and iii) junctional tension v row number varying the cell-substrate adhesion ($h_0$) parameter (a) and the contractility ($K$) parameter (b). For these simulations $J = 0.375, s = 0.001, c = 0.4x10.^2$

**Supplementary Figure 1.** Relationship between the area and the side length square for different regular polygons.

**Supplementary Figure 2.** a) Calculations of offset between apical and basal areas of cells (Total and in the directions orthogonal and parallel to the island edge) along the island radial direction. For these simulations $K = 0.45, J = 0.375, ho = 0.01, s = 0.001, c = 0.4x10^2$. b and c) Calculations of offset between apical and basal areas of cells (along the island radial direction) for different values of cell motility (b) and stiffness (c). For these simulations $K = 0.45, J = 0.375, ho = 0.01, s = 0.001, c = 0.4x10^2$ For the simulations in (b) $K = 0.45, J = 0.375, ho = 0.005, 0.01, 0,02, s = 0.001, c = 4x10^2$. d) Plots of junctional tension across Islands before and after increase contractility in 4 cells in the center of the island by 20%. Parameters for these simulations are $K = 0.45$ (for the entire island), $K = 0.54$ (for cells with increased contractility), $J = 0.375, ho = 0.01, s = 0.001$ and $c = 4x10^2$.

**Supplementary Movie 1.** Simulation of an expanding epithelial island. At t=0, the island is confined to a square and cells have an average area smaller than their preferred area ($\zeta < 1$). Then the barrier is removed and cells are allowed to expand and acquire its equilibrium shape. In blue is indicated the cell-substrate interface area, in red is the apical area of cells and in green is the direction of the longer cell axis orientation. The pseudo-color pink on the apical area of cells indicates its aspect ratio. Note in the simulation that once the onset of the island expansion occurs, the vectors of the cell long axis orients perpendicular to the island edge and this orientation is propagated to cells inside the island.

**Figures**

Figure 1.

A

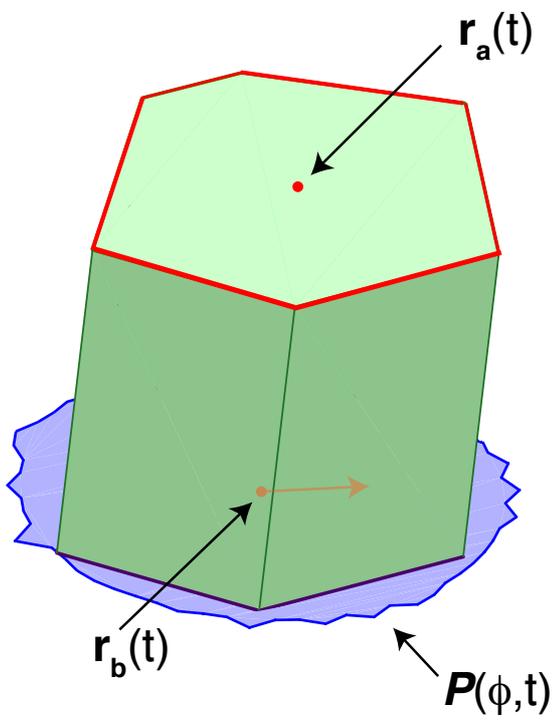

$r_a(t)$

$r_b(t)$

$P(\phi, t)$

B

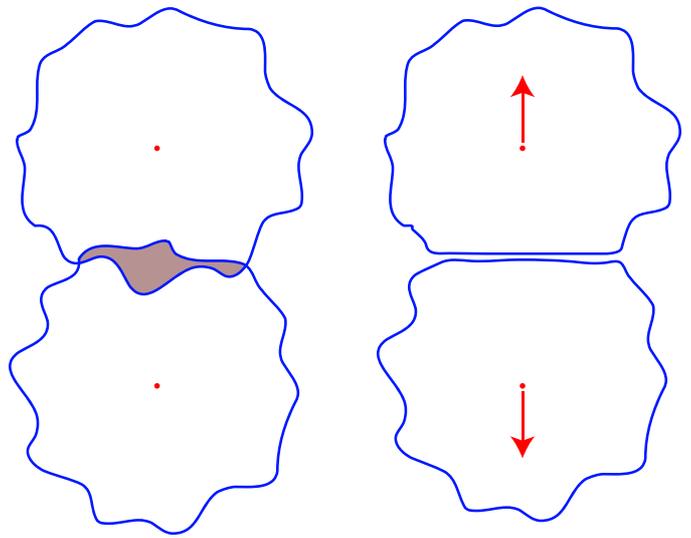

C

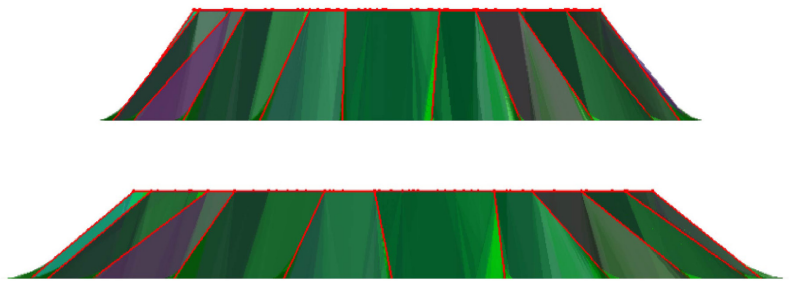

# Figure 2

**A** 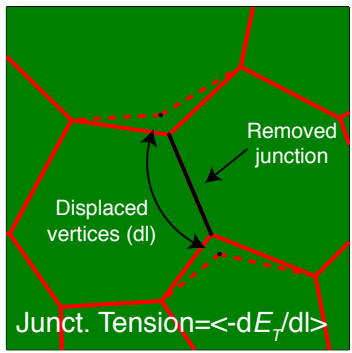

**B** 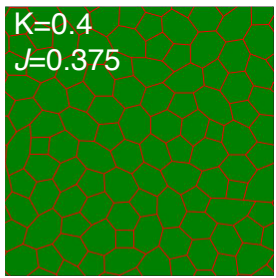 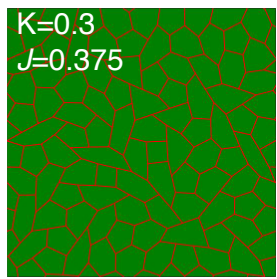 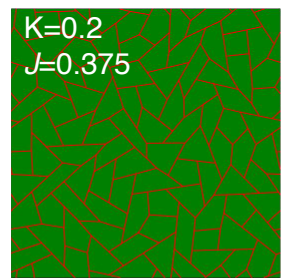

**C** 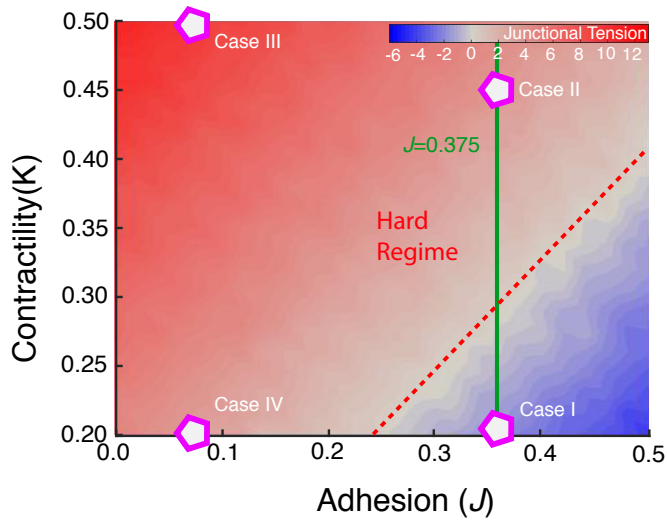

**D** 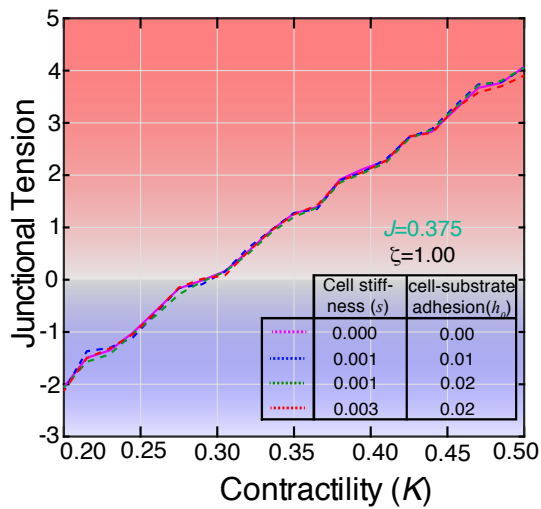

# Figure 3.

## A

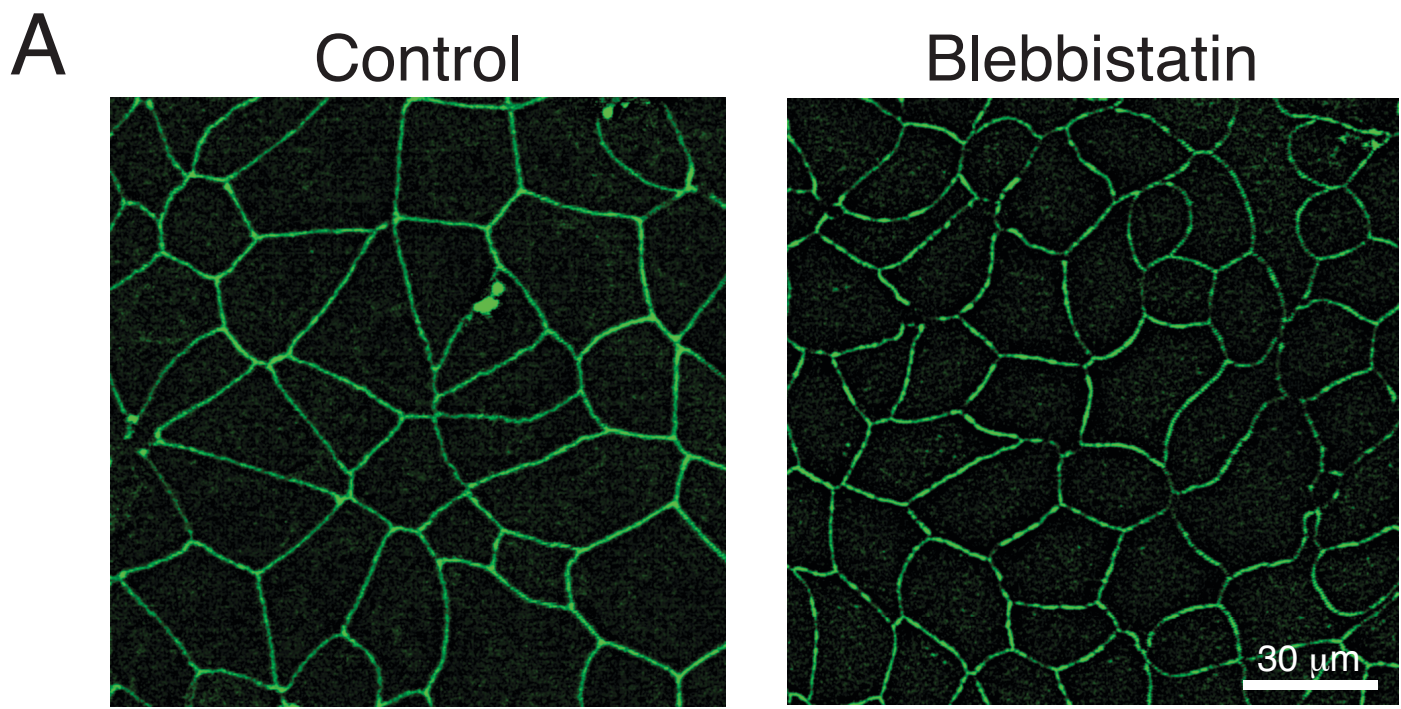

Control | Blebbistatin

## B

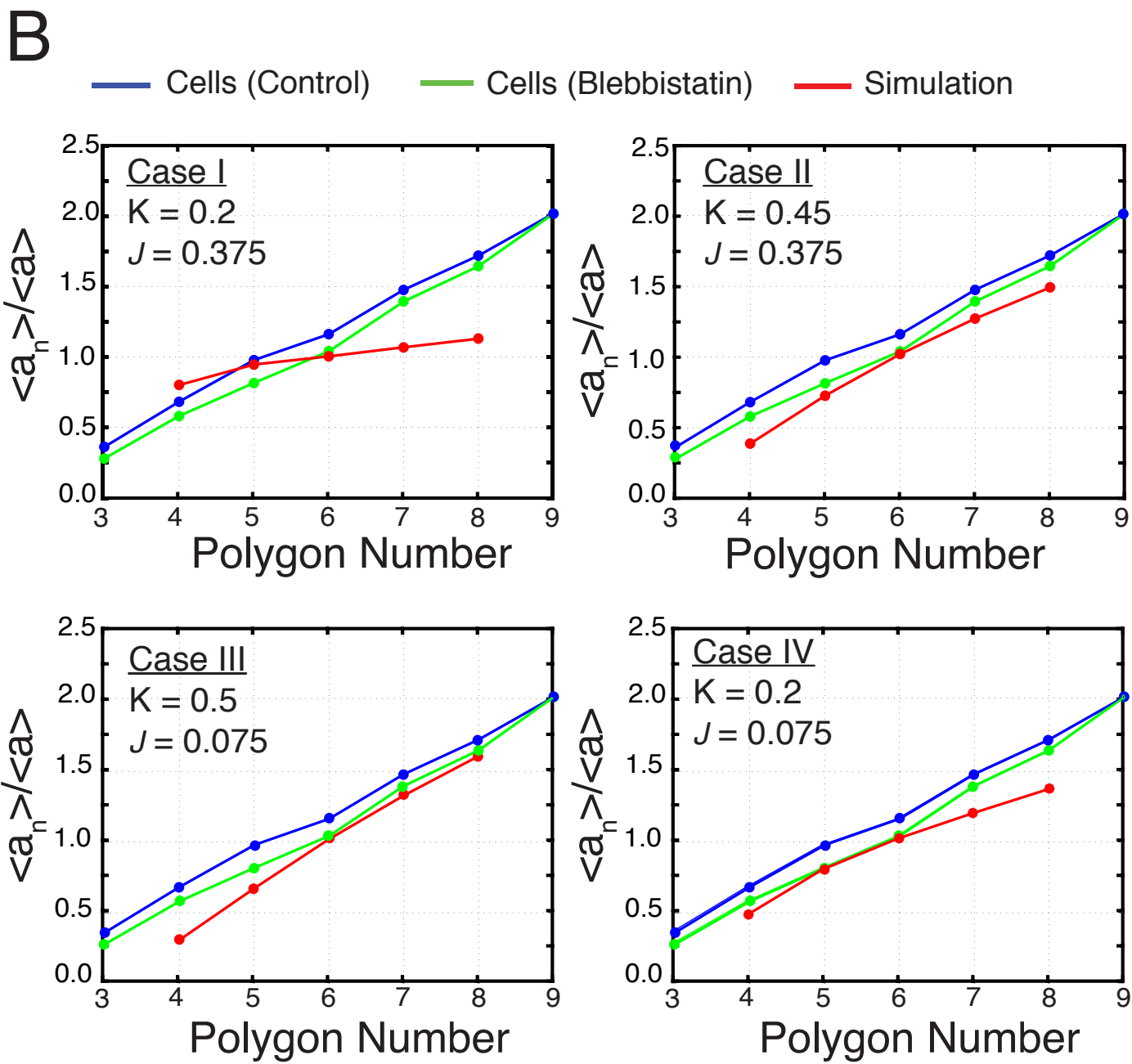

— Cells (Control)    — Cells (Blebbistatin)    — Simulation

Case I: K = 0.2, $J$ = 0.375

Case II: K = 0.45, $J$ = 0.375

Case III: K = 0.5, $J$ = 0.075

Case IV: K = 0.2, $J$ = 0.075

Axes: $\langle a_n \rangle / \langle a \rangle$ vs Polygon Number

# Figure 4.

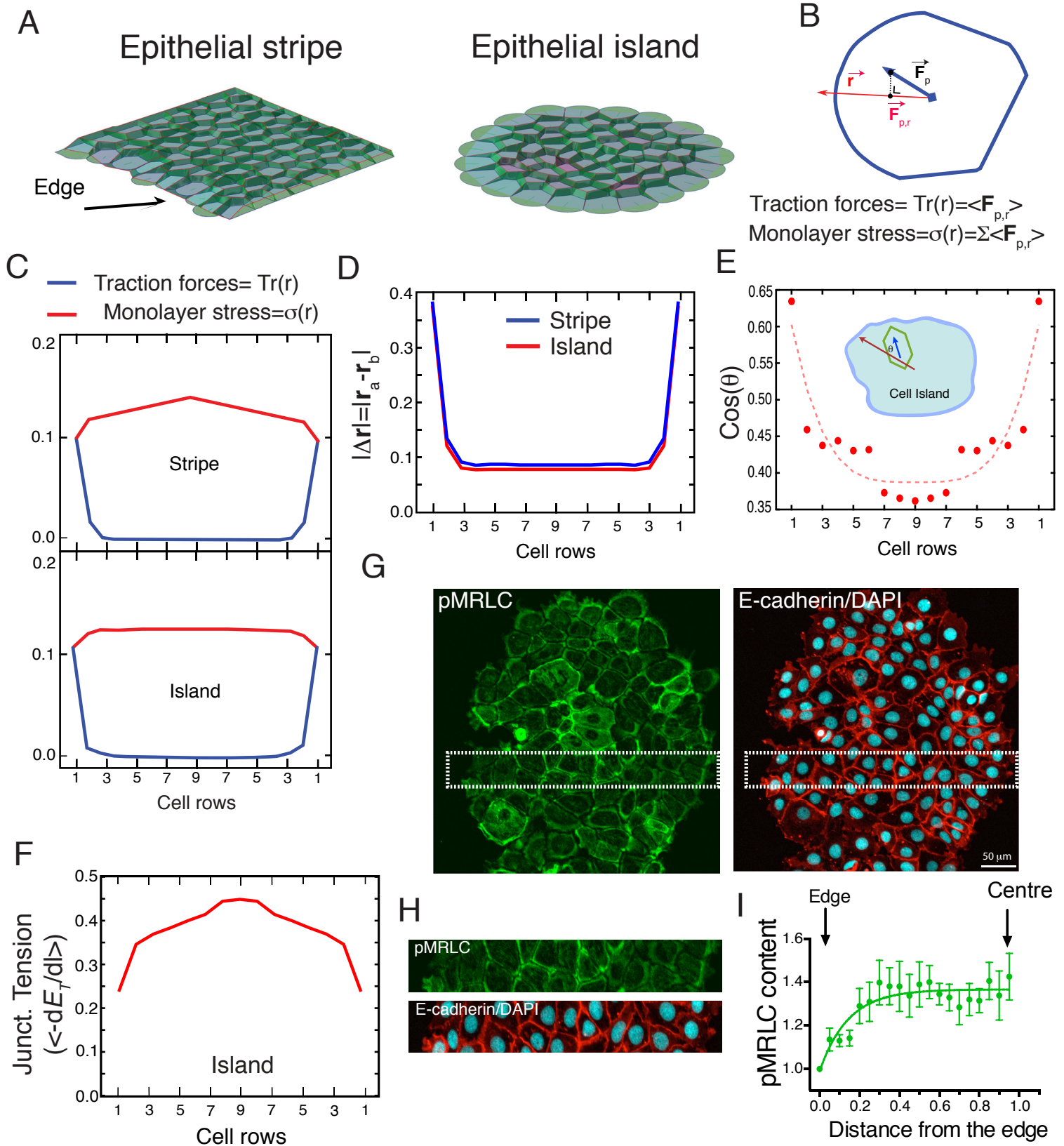

# Figure 5.

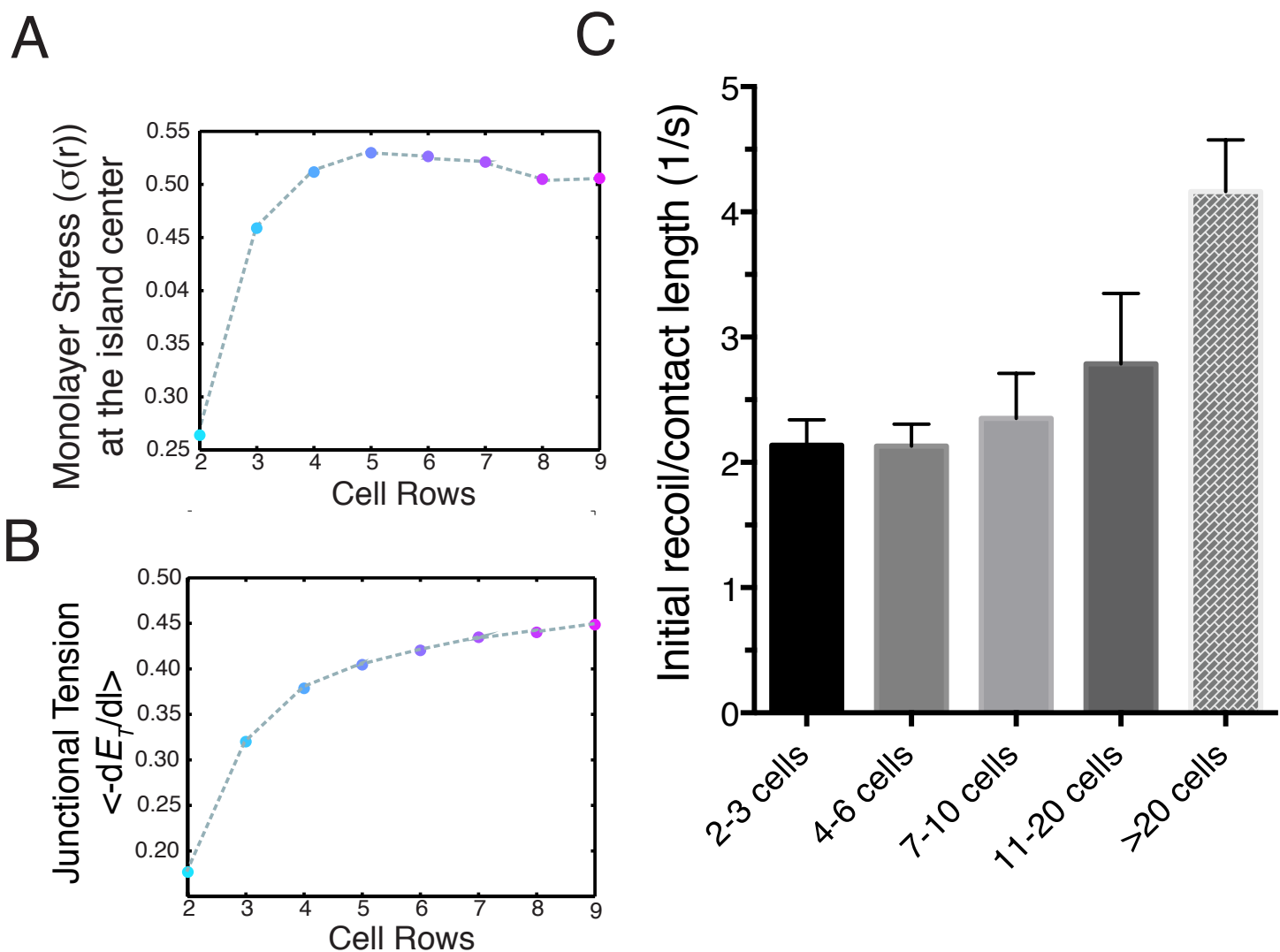

# Figure 6.

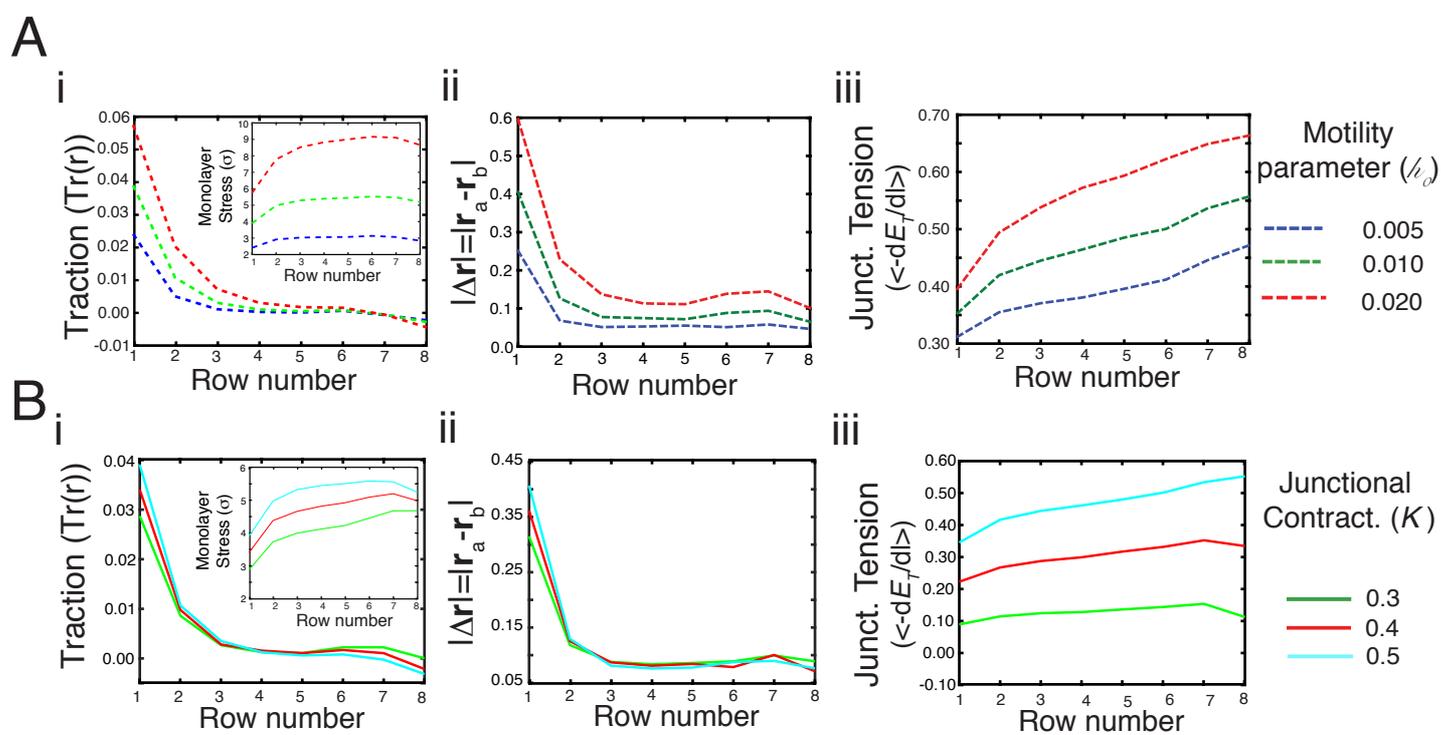

## Supplementary Figure 1.

| Polygon class | | Area= $\eta*$(side length)$^2$ |
|---|---|---|
| Triangle | 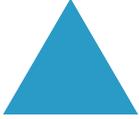 | $\eta=0.433$ |
| Square | 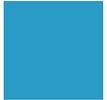 | $\eta=1.000$ |
| Pentagon | 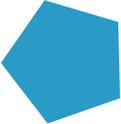 | $\eta=1.720$ |
| Hexagon | 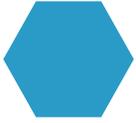 | $\eta=2.598$ |

# Supplementary Figure 2.

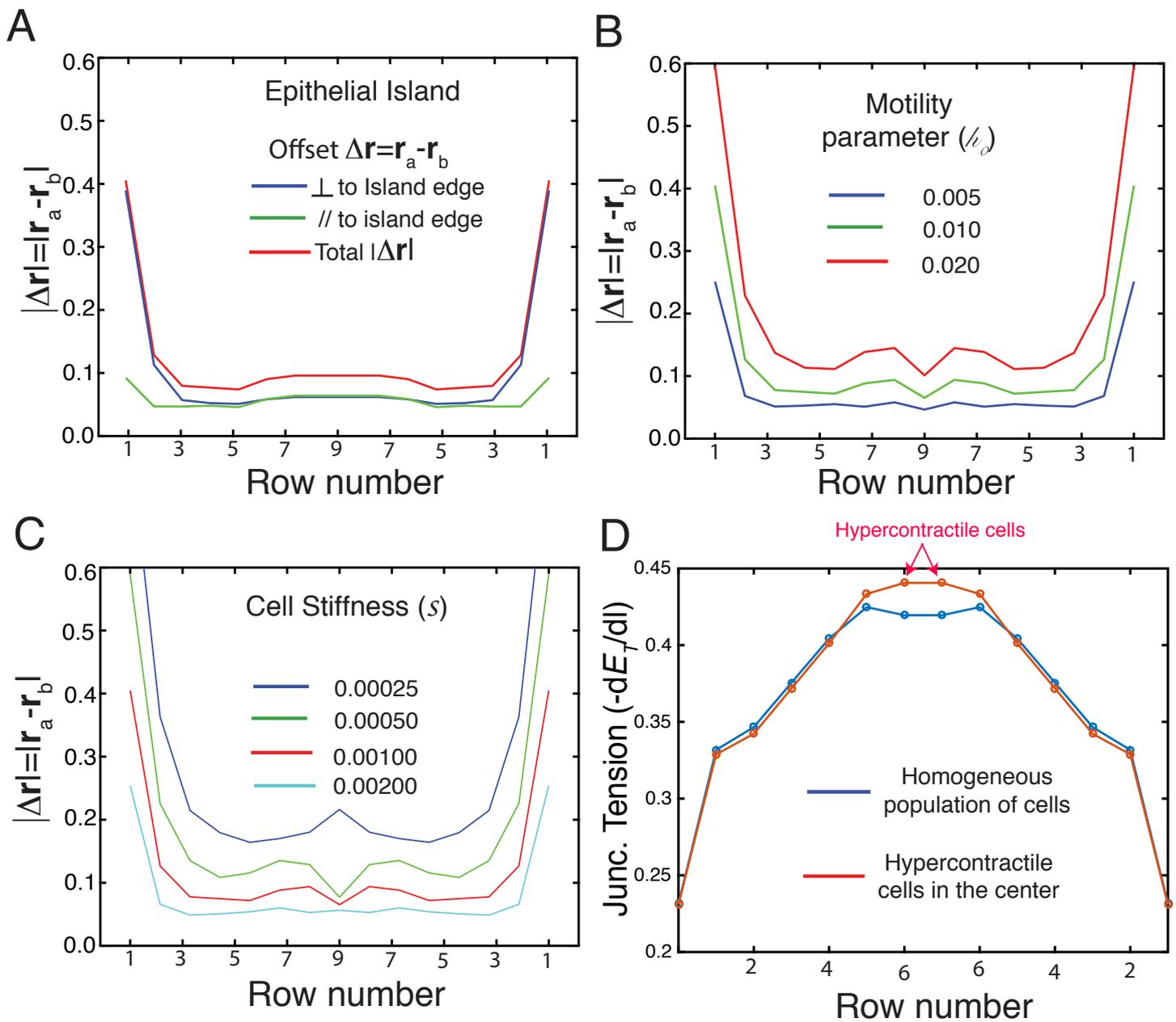